\documentclass[12pt,aps,amsmath,latexsym,amsfonts]{JHEP3}

\usepackage{epsfig}
\usepackage{amsfonts,amssymb,amsmath}
\usepackage[hang]{subfigure}

\newcommand{\be}{\begin{equation}}
\newcommand{\ee}{\end{equation}}
\newcommand{\bea}{\begin{eqnarray}}
\newcommand{\eea}{\end{eqnarray}}

\title{The Cosmology of Asymmetric Brane Modified Gravity}
\author{Eimear O'Callaghan$^1$\thanks{Email: e.e.o'callaghan@durham.ac.uk} , 
Ruth Gregory$^1$\thanks{Email: r.a.w.gregory@durham.ac.uk} , 
Alkistis Pourtsidou$^2$\thanks{Email: ppxap1@nottingham.ac.uk}
\\
~$^1${\it Institute for Particle Physics Phenomenology, 
Durham University, South Road, Durham, DH1 3LE, UK}\\
~$^2${\it School of Physics and Astronomy, University of Nottingham,
University Park, Nottingham, NG7 2RD, UK}}
\abstract{
We consider the asymmetric branes model of modified gravity, which 
can produce late time acceleration of the universe and compare 
the cosmology of this model to the standard $\Lambda$CDM model 
and to the DGP braneworld model. 
We show how the asymmetric cosmology at relevant physical scales can
be regarded as a one-parameter extension of the DGP model, and
investigate the effect of this additional parameter on the expansion
history of the universe. 
}

\preprint{IPPP/09/35, DCPT-09/70}

\begin{document}
\newcommand{\zed}{$\mathbb{Z}_2$}

\section{Introduction}

Recent observations of high redshift supernovae suggest that 
the universe is currently undergoing a phase of 
acceleration \cite{Riess1,Perl}. 
This is usually explained by the presence of `dark energy', 
some as yet unknown negative pressure fluid. In 
the standard model of cosmology, the $\Lambda$CDM model,
this dark energy is assumed to be vacuum energy in the 
form of a small, positive cosmological constant. 
Measurements of the cosmic microwave background (CMB) 
anisotropies by WMAP \cite{WMAP5} and large scale 
structure surveys \cite{Cole:2005sx} suggest that this fluid 
must make up $\sim 70\%$ of the content of the universe,
the remainder being made up of matter, both baryonic 
matter ($\sim 4\%$) and dark matter.  Despite the fact 
that the $\Lambda$CDM model is currently the best fit to the cosmological
data, there are theoretical issues with the cosmological constant. 
For a naive estimate of $\Lambda$, 
there is a difference of around 120 orders of magnitude
between the theoretical prediction and the value inferred 
from observations.  As yet, no satisfactory explanation
has been put forward to resolve this discrepancy and 
explain the observed smallness of $\Lambda$.

Rather than trying to explain why the cosmological constant is 
so small, various alternative mechanisms for achieving late 
time acceleration have been explored. These can essentially be
divided into models which modify the matter content of the 
universe, such as quintessence \cite{quint}, or the more 
phenomenological Cardassian \cite{card}, and Hobbit models \cite{hobbit},
and those which modify the gravitational interaction, 
such as MOND \cite{MOND}, $f(R)$ theories \cite{fofR}, and braneworld models.
The braneworld scenario is a set-up in which we have extra dimensions 
in nature, which are hidden because we are confined to live on a 
slice - a {\it brane} in the higher dimensional spacetime. The most 
common models, ADD \cite{ADD} and Randall-Sundrum (RS) \cite{RS} , 
lead to an effective theory of braneworld gravity which is Einstein gravity at
large scales, but with small scale Kaluza-Klein modifications. 
However, very early on in braneworld research, it was realised 
that braneworlds could also display large scale modifications 
of gravity \cite{mill,GRS,DGP,VS}. The DGP model in 
particular \cite{DGP,DGPcos} has received a great deal of 
attention as a possible viable cosmological alternative
to $\Lambda$CDM. However, while being attractive from the 
phenomenological point of view, the DGP model has 
inconsistencies, such as ghosts \cite{GHOST,CGKP}, pressure
singularities and tunnelling instabilities \cite{DGPsing,GKMP}
(although see \cite{DG} for counter-arguments).
As such, it is surprising that so much focus has centered on
DGP in comparison to other braneworld modified gravities.

DGP, like many braneworld models, has a $\mathbb{Z}_2$ symmetry 
around the braneworld, but interestingly if one relaxes this 
symmetry, it is also possible to get IR modifications of gravity such 
as the asymmetric models of Padilla, \cite{Padillovicz,tonyboy}, 
and the hybrid asymmetric-DGP type ``stealth" model of Charmousis, 
Gregory and Padilla \cite{CGP} (see also \cite{SSST}). In each of these 
models, the cosmological constant {\it and} the Planck masses are
different on each side of the brane. In the asymmetric 
model (which we focus on here) there is a strong hierarchy 
between the Planck masses and the adS curvature scales on each
side of the brane.
On one side of the brane, there is a large cosmological 
constant and Planck mass and the interior of the bulk is 
retained, on the other side of the brane the cosmological constant 
and Planck mass are low, and the exterior of the bulk is kept.  
Keeping the bulk interior produces a localizing effect on the 
braneworld gravity, whereas keeping the exterior tends to produce 
an opening up effect and modifies gravity in the infrared \cite{KK}.  
With a judicious choice of scales,  it is possible to have a regime in
which gravity is effectively 4D, before opening up at very large scales. 
(At small scales of course, the KK modes cause gravity to be 
effectively 5D.) In \cite{Padillovicz,tonyboy} this model has 
been extensively tested from the point of view of
particle physics, but it is less clear that it will pass 
cosmological tests and in particular, reduce to standard 
4-dimensional General Relativity at early times.

In this paper we explore the cosmology of the asymmetric (AC) model, 
focussing in particular on the type Ia supernova data 
\cite{DavisWV}, and the expansion parameters from WMAP 
\cite{WMAP5,WMAPcos}. We first show how the AC model
can be viewed as a one parameter extension of the DGP model over 
a wide range of scales.
We then explore the effect of this additional parameter
on the expansion history of the universe, making reference to the
Supernova and WMAP data. Finally, we discuss how the additional
parameter makes it more difficult to simultaneously fit
the various observational constraints.
We also comment on the inclusion of a bulk black hole.

\section{Asymmetric Braneworld Models}

We start by reviewing the asymmetric branes model, and 
deriving the cosmological equations. The action can be written as
\cite{tonyboy} 
\begin{equation}\label{1}
S = S_{bulk} + S_{brane}
\end{equation} 
where the `bulk' action contains the gravitational field dynamics,
and is given by the Einstein Hilbert and Gibbons Hawking terms:
\begin{equation}\label{2}
S_{bulk} = \sum_{i=1,2}M_{i}^{3}\int_{\mathcal{M}_{i}}d^5x\sqrt{-g}(R -
2\Lambda_{i})+2M_{i}^{3}\int_{brane}d^4x\sqrt{-\gamma}K^{(i)}.
\end{equation} 
Here, $g_{ab}$ is the bulk metric with corresponding Ricci scalar $R$. The
metric induced on the brane is 
\begin{equation}\label{3}
\gamma_{ab}=g_{ab}-n_{a}n_{b}
\end{equation} 
where $n^a$ is the unit normal to the brane in $\mathcal{M}_{i}$ 
pointing out of $\mathcal{M}_{i}$. The extrinsic curvature of 
the brane in $\mathcal{M}_{i}$ is given by 
\begin{equation}\label{4}
K_{ab}^{(i)} = \gamma_{a}^{c}\gamma_{b}^{d}\nabla_{(c}n_{d)}. 
\end{equation} 
The brane action for the asymmetric branes model is given by 
\begin{equation}\label{5}
S_{brane} =  \int_{brane}d^4x(-\sigma\sqrt{-\gamma} + \mathcal{L}_{matter}) 
\end{equation} 
where $\sigma$ is the brane tension and $\mathcal{L}_{matter}$ describes the
matter content on the brane.

Note that with braneworld models, there is a subtlety in how we encode
the gravitational action. We can either view the brane as a genuine zero
thickness object -- a mathematical boundary between two different spacetimes
(which just happen to be mirror images of each other in the usual
\zed\ braneworlds, such as Randall-Sundrum) -- or as the zero thickness
limit of some finite size object, an approximation to the domain wall
of the early braneworld models \cite{EBW}. The equivalence
of these two descriptions has been well established \cite {Israel}, 
as well as an understanding of the next to leading order corrections
\cite{GG}. (Although see \cite{GTV} for an interesting
discussion of possible cosmological consequences of finite width.)
These different physical perspectives, boundary vs.\ $\delta$-function,
translate into a different way of expressing the action, i.e.\ whether or
not we use the Gibbons Hawking term. In the asymmetric case, since 
the Planck mass is different on each side of the brane, the 
boundary description is somewhat more natural, and
makes it easier to obtain the correct equations of motion.

The equations of motion for each bulk are given by the Einstein equations 
\begin{equation} \label{6}
R_{ab} -\frac{1}{2}Rg_{ab} = -\Lambda_i g_{ab}
\end{equation}
while the brane equations of motion are found from the Israel conditions 
\cite{Israel} to be
\begin{equation}\label{7}
2\langle M^3K_{ab}\rangle-\frac{\sigma}{6}\gamma_{ab}
=\frac{1}{2}\Big(T_{ab}-\frac{1}{3}T\gamma_{ab}\Big),
\end{equation} 
obtained by varying (\ref{1}) with respect to the induced brane metric
$\gamma_{ab}$. The energy-momentum tensor for the 
additional matter on the brane is given by 
\begin{equation} \label{8}
T_{ab} =
-\frac{2}{\sqrt{-\gamma}}\frac{\partial\mathcal{L}_{m}}{\partial\gamma^{ab}}.
\end{equation}

The background metric $\bar{g}_{ab}$ is found by solving the 
equations of motion with $T_{ab} = 0$, and may be written as 
\begin{equation} \label{bkmet}
ds^{2}=\bar{g}_{ab}dx^{a}dx^{b}=a^{2}(y)\eta_{\mu\nu}dx^{\mu}dx^{\nu} + dy^{2} 
\end{equation}
where $x^a = (x^{\mu},y)$ with the brane at $y = 0$, and $a(y)$ is the
warp factor which has the general form:
\begin{equation}
a_i(y)=e^{-\theta_i k_i |y|}
\end{equation}
where $\theta_i=\pm1$, the subscript $i=1,2$ refers to the two sides
of the brane ($i=1$ being $y<0$), and $\Lambda_i = -6k_i^2$ 
defines the adS curvature scale on each side. 

The metric (\ref{bkmet}) and the equations of motion also impose a 
condition on the brane tension: 
\begin{equation} \label{10}
2 \langle M_i^3\theta_i k_i \rangle = \frac{\sigma}{6}
\end{equation} 
where $\langle Z\rangle = (Z_{1}+Z_{2})/2$ and $\Delta Z = Z_{1}-Z_{2}$
for a quantity $Z_{i}$ differing across the brane.

Three separate cases of this model were considered in \cite{tonyboy}
for different $\theta_{i}$ values: (i) the Randall-Sundrum (RS) case, 
$\theta_{1} = \theta_{2} = 1$,
(ii) the inverse Randall-Sundrum case, $\theta_{2}=\theta_{1}=-1$, and
(iii) the mixed case, where $\theta_{1}=-\theta_{2}=1$.
If $\theta_{1}$ corresponds to the left-hand side of the brane $(y<0)$ and
$\theta_{2}$ corresponds to the right-hand side of the brane $(y>0)$, then the
RS case has the warp factor decaying away from the brane on both sides, while
the inverse RS case has the warp factor growing on both sides. In the mixed
case, the warp factor decays away from the brane on the left, 
whilst growing on the right. As explained in detail in \cite{tonyboy,KK},
whereas 4-dimensional Einstein gravity cannot be reproduced at any 
scale in the inverse RS case, it can be achieved in the RS and 
mixed cases, along with infra-red (IR) modifications.  However, only 
the mixed case will approach a de Sitter state at late times, leading 
to exponential late-time acceleration without an effective cosmological
constant \cite{Padillovicz}. Therefore, only the mixed case where
$\theta_{1}=-\theta_{2}=1$ is considered from now on.

Turning to cosmological solutions, since we have Einstein gravity in the
bulk, we know that the bulk is completely specified by
the AdS-Schwarzschild metric  \cite{BCG}
\begin{equation}\label{13}
-h_i(r)dt^2+\frac{dr^2}{h_i(r)}+r^2d\textbf{x}_{\kappa}^2
\end{equation}
where
\begin{equation}\label{hgen}
h_i(r)=r^2k_i^2 + \kappa - \frac{\mu_i}{r^2} \; .
\end{equation}
For simplicity, we will take the case where there is no black hole in either
bulk, $\mu_i=0$.

In order to construct the brane, we glue a solution in $\mathcal{M}_1$ 
to a solution in $\mathcal{M}_2$, where the brane will form the 
common boundary. Then, in $\mathcal{M}_i$, the boundary 
$\partial\mathcal{M}_i$ is given by the section ($t_i(\tau)$, 
$a_i(\tau)$, $\textbf{x}^{\mu}$) of the bulk metric (\ref{13}), 
where $\tau$ is the proper time of an observer comoving with the 
boundary, so that
\begin{equation}\label{15}
-h_i(a_i)\dot{t}^2+\frac{{\dot{a}_i}^2}{h_i(a_i)}=-1
\end{equation}
where the differentiation is with respect to $\tau$. The 
outward pointing unit normal to $\partial\mathcal{M}_i$ is now given by
\begin{equation}\label{16}
n_a= \theta_i(-\dot{a}_i(\tau),\dot{t}_i(\tau),\textbf{0})
\end{equation}
where $\theta_i=\pm1$ as before. For $\theta_i=1$, $\mathcal{M}_i$ 
corresponds to $0\leq a< a_i(\tau)$, while for $\theta_i=-1$, 
$\mathcal{M}_i$ corresponds to $a_i(\tau)<a<\infty$. The induced 
metric on $\partial\mathcal{M}_i$ is that of a FRW universe,
\begin{equation}\label{17}
ds^2=-d\tau^2+a_i^2d\textbf{x}_{\kappa}^2.
\end{equation}
Since the brane coincides with both boundaries, the metric on the 
brane is only well defined when $a_1(\tau)=a_2(\tau)=a(\tau)$ and 
the Hubble parameter is then defined as $H=\frac{\dot{a}}{a}$. If 
we now introduce a homogeneous and isotropic fluid on the brane, 
whose energy-momentum tensor is given by \cite{Padillovicz}
\begin{equation}\label{18}
T_{ab}=(\rho+p)\tau_a\tau_b+p\gamma_{ab},
\end{equation}
with energy density $\rho$, pressure $p$ and $\tau^a$, the velocity 
of a comoving observer (which in $\mathcal{M}_i$ is 
$\tau^a=(\dot{t}_i(\tau),\dot{a}(\tau),\textbf{0})$), we can 
evaluate the spatial components of (\ref{7}). 
Doing this and using (\ref{15}) to substitute for $\dot{t}$, we find 
\begin{equation}\label{19}
2\Bigg\langle M_i^3\theta_i \sqrt{H^2+\frac{h}{a^2}}
\Bigg\rangle =\frac{(\rho+\sigma)}{6}.
\end{equation}
Substituting for $\sigma$ using (\ref{10}), and $h(a)$ using (\ref{hgen})
with $\mu_i=0$, the modified Friedmann equation for the mixed case is
\begin{equation}\label{fried}
\rho =
6\bigg[M_1^3\Big(\sqrt{H^2 + \frac{\kappa}{a^2} +k_1^2}-k_1\Big)
-M_2^3\Big(\sqrt{H^2 + \frac{\kappa}{a^2} +k_2^2} -k_2\Big)\bigg].
\end{equation}

From \cite{tonyboy,KK}, we know that there is a range of scales over
which gravity is four dimensional, given by
\be
k_1^{-1} \ll r \ll r_c = \frac{M_1^3}{M_2^3 k_1}
\ee
which clearly requires $M_1\gg M_2$. 
For this model to be phenomenologically viable, this range of scales 
must be appropriate. Since we are looking at $r_c$ as representing
the scale at which late time acceleration sets in, we expect
the crossover scale to be of order the current horizon size, 
$r_c\sim H_0^{-1}$.  On the other hand, table-top tests of General 
Relativity \cite{Kapner} have confirmed its validity down to sub-mm 
scales, which fixes our largest frequency scale, $k_1$ 
(the UV cut-off of the theory), so that $\frac{1}{k_1}\sim 0.1$mm. 
These constraints give us a large hierarchy of scales, and, as already
noted, require a large hierarchy in the parameters.

It is interesting to see these scales emerge from an analysis of
the Friedmann equation (\ref{fried}). Obviously (\ref{fried})
looks nothing like the standard Friedmann equation, and so can only reduce
to such in certain asymptotic limits. Setting $\kappa=0$ for simplicity,
and using
\be
\sqrt{H^2 + k^2} \simeq 
\begin{cases}
k + \frac{H^2}{2k} & {\rm for}\; H\ll k\cr
H & {\rm for}\; H \gg k \cr
\end{cases}
\ee
we see that we can only get the $H^2$ behaviour required if $k_1 \gg k_2$,
and we take $H\ll k_1$. 
In this r\'egime, the Friedmann equation can be written as
\be
\label{asyfrweq}
\rho \simeq 
3 \frac{M_1^3}{k_1} H^2 - 6 M_2^3 \Big(\sqrt{H^2+k_2^2} -k_2\Big)
\ee
We therefore see the existence of an accelerating vacuum whenever
\be
H_A^2 = 4 k_1 \frac{M_2^3}{M_1^3} \left ( k_1 \frac{M_2^3}{M_1^3} - 
k_2 \right ) > 0
\label{betalim}
\ee
We can also read off the 4D Planck mass $m_{pl}^2 = 1/8\pi G$
by comparing with the standard 4D Friedmann equation:
\begin{equation}\label{24}
\rho=3m_{pl}^2H^2 
\end{equation}
as
\begin{equation}\label{planckmass}
m_{pl}^2\simeq\frac{M_1^3}{k_1}>0
\end{equation}
This agrees with the expression derived in \cite{tonyboy}, and also
with a direct computation of the propagator (see appendix \ref{appA}).

We would like to compare (\ref{asyfrweq}) with the cosmological
equations from the DGP model. The DGP model is characterized by
an induced curvature term on the brane, and (in its original form) is
\zed\ symmetric around the brane, which is tensionless and embedded
in 5D Minkowski space \cite{DGP}:
\begin{equation}\label{34} 
S_{\rm DGP} =M_{5}^3\int d^5X\sqrt{G}R_{(5)} + M_4^2\int d^4x\sqrt{|g|}R_{(4)}
\end{equation}
The brane cosmological equations from this action are given by \cite{DGPcos}
\be
\rho = 6 M_4^2 H^2 \mp 12 M_5^3 H 
\label{DGPeq}
\ee
The choice of sign in the linear Hubble term is due to the choice of
which part of the bulk is kept. The minus sign, corresponding to the
exterior being kept, is the {\it self-accelerating} branch, which has
late time cosmological acceleration. The crossover 
scale $r_{DGP} = M_4^2/2M_5^3$
corresponds to the scale at which gravity ceases to be 4D, and the
extra dimension opens up. 

To compare the asymmetric and DGP models, note that if we take $H\gg k_2$, 
then we may approximate the second bracket in (\ref{asyfrweq}), and obtain
\be
\rho \simeq 
3 \frac{M_1^3}{k_1} H^2 - 6 M_2^3 H 
\ee
which is of course (\ref{DGPeq}) after suitable substitution.

Over a large range of scales therefore, AC cosmology can be viewed
as a generalization of DGP cosmology. To parametrize this in a simple
way for our analysis, we set
\begin{equation}
\label{reparam}
\alpha = \frac{k_1}{H_0}\frac{M_2^3}{M_1^3} \qquad  
\beta = \frac{k_2}{H_0}\qquad
E = \frac{H}{H_0}
\end{equation}
where $H_0$ is the current value of the Hubble parameter. 
(\ref{fried}) then becomes:
\be
\label{abfried}
\rho = 3 m_{pl}^2 H_0^2 \left [ E^2 + \frac{\kappa}{a^2H_0^2}- 2\alpha \left (
\sqrt{E^2+ \frac{\kappa}{a^2H_0^2} + \beta^2} - \beta \right ) \right].
\ee
Here, $\alpha = (2H_0r_{DGP})^{-1}$ is essentially the same as the 
DGP crossover scale, and $\beta$ is the new parameter coming from 
the asymmetric physics. It is precisely the effect of this new 
parameter which we seek to explore.

\section{Asymmetric Cosmology}

In order to explore the effect of the AC model, it is
useful to rewrite the Friedmann equation in an Einstein form 
by solving (\ref{abfried}) for $E=H/H_0$ :
\begin{equation}\label{asycos}
E(z)^2 = \Omega_k (1+z)^2 + \Omega_i(1+z)^{3(1+w_i)} + 2\alpha(\alpha-\beta) 
+ 2\alpha \sqrt{(\alpha-\beta)^2+\Omega_i(1+z)^{3(1+w_i)}}. 
\end{equation}
Here, an implicit sum over the various contributions to the energy density
with equations of state $p_i = w_i \rho_i$ is understood and
$\Omega_k = - \kappa/a_0^2H_0^2$.
Note that the $+$ root of the quadratic is required to get the correct
$\Omega \to 0$ limit of the Israel equations.
We can now readily compare the AC model
with $\Lambda$CDM and DGP, which are 
implicitly contained in (\ref{asycos}):
$\alpha=0$ and we include an $\Omega_\Lambda$ for $\Lambda$CDM,
and $\beta=0$ for DGP. 
Since the DGP model has been carefully analysed with cosmological
expansion data (see e.g.\ \cite{DGPcomp}), here we focus qualitatively
on the additional features the $\beta$-term brings relative to DGP.

The aim of gravitationally driven late time acceleration is to
avoid using a cosmological constant ($\Omega_\Lambda$), therefore 
evaluating (\ref{asycos}) at the current time gives a constraint
between the model parameters $\alpha,\beta$ and the current matter 
and curvature densities:
\begin{equation}\label{30}
\Omega_{m}= 1-\Omega_k - 2\alpha(\sqrt{1- \Omega_k +\beta^2}-\beta)
\end{equation}
(ignoring the relatively insignificant radiation component). 
This means that once $\Omega_m$ and $\Omega_k$ are fixed, 
the asymmetric cosmology 
forms a one parameter family of solutions (note that DGP is entirely 
constrained by fixing $\Omega_m$ and $\Omega_k$).
From (\ref{reparam}), we see that both $\alpha$ and $\beta$ are positive, and 
in addition self acceleration requires $\beta<\alpha$ from (\ref{betalim}).
Thus our additional parametric degree of freedom in the asymmetric model
has a fairly limited range. 

The modified Friedmann equation (\ref{asycos}) shows clearly the effect
of $\beta$ over the range $[0,\alpha]$. As already noted, $\beta=0$
corresponds to the DGP model, with $\alpha^2 = \Omega_{r_c}$ in
the usual notation of encoding the DGP crossover scale as an effective
DGP $\Omega$ contribution. The other limit, $\beta=\alpha$, corresponds 
to an $n=1/2$ Cardassian model \cite{card}, or, alternatively,
a dark energy fluid with (constant) equation of state $w = -1/2$.

As with DGP, relaxing the constraint of flatness leads to a wider
range of parameter choice:
\be
\frac{(1-\Omega_k-\Omega_m)}{2\sqrt{\Omega_m}} \geq \alpha \geq
\frac{(1-\Omega_k-\Omega_m)}{2\sqrt{1-\Omega_k}}
\ee
and the cosmological models now form a three parameter family, which
can be labelled using $\{\Omega_m,\Omega_k,\alpha^2\}$ (or by trading
one of the $\Omega$ parameters for $\beta$). In order to more readily
compare with DGP results, we will use the former parametrization,
and compare results in the $\{\Omega_m, \alpha^2\}$
plane (recall $\alpha^2 = \Omega_{r_c}$ in DGP) either 
for various fixed $\beta$ values with $\Omega_k$ varying, or fixed
$\Omega_k$ values with $\beta$ varying.
In spite of the enlarged parameter space, the asymmetric cosmology 
turns out to be under more cosmological tension than DGP. 

A nice way to encode this information is to consider the 
effective dark energy which is the difference between the square of the
Hubble parameter and the matter content \cite{Varun}:
\begin{equation}\label{effDE}
\Omega_{DE}(z) = 2\alpha(\alpha-\beta) + 2\alpha \sqrt{(\alpha-\beta)^2
+\Omega_i(1+z)^{3(1+w_i)}} \;,
\end{equation}
where we have taken $\Omega_k=0$ for simplicity. The
effective dark energy pressure is then the discrepancy between
the Einstein pressure and the actual pressure:
\begin{equation}\label{effT}
\Pi_{DE}(z) = - \alpha \left [ 2(\alpha-\beta) 
+ \frac{2(\alpha-\beta)^2 + (1-w_i)\Omega_i(1+z)^{3(1+w_i)}}
{\sqrt{(\alpha-\beta)^2+\Omega_i(1+z)^{3(1+w_i)}}} \right ].
\end{equation}
Using these, we can find an effective equation of state,
\be
w_{DE}(z) = - 1 + \frac{(1+w_i)\Omega_i(1+z)^{3(1+w_i)}}
{2[(\alpha-\beta)\sqrt{(\alpha-\beta)^2+\Omega_i(1+z)^{3(1+w_i)}}
+ (\alpha-\beta)^2+\Omega_i(1+z)^{3(1+w_i)}]}.
\ee
This shows how the equation of state always has $w\geq-1$ for
$\alpha\geq\beta$, and thus the model can never enter a phantom regime.
We also see how for $\beta>0$, $w$ is raised from its DGP ($\beta=0$) 
value (see figure \ref{fig:eos}). 
Overall therefore, we expect that expansion data will favour a lower
$\Omega_m$ in both DGP and AC models.
\FIGURE{\label{fig:eos}
\includegraphics[width=10cm]{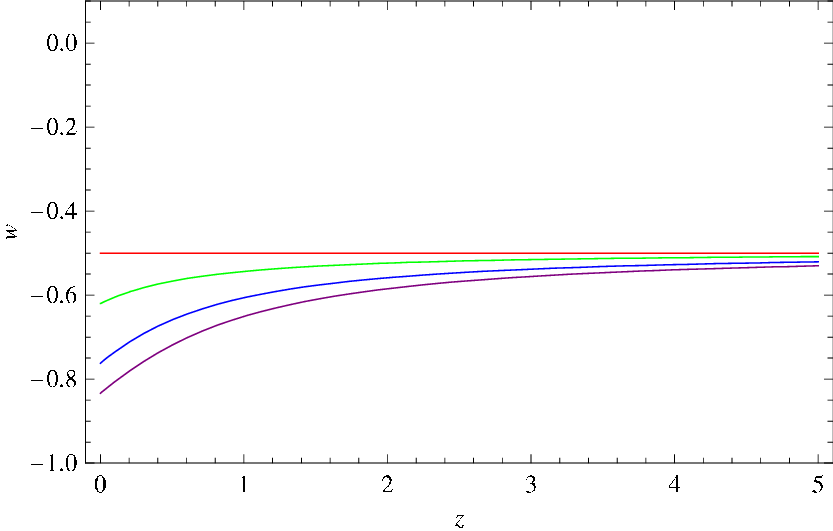}
\caption{Variation of the effective dark energy equation of
state taking $\Omega_{m}=0.25$. 
The values of $\alpha$ run from the minimum allowed ($\beta=0$, the DGP value)
in purple, to the maximum value, $\alpha=\beta$, shown in red.
These clearly show how increasing $\beta$ neutralizes acceleration 
in asymmetric cosmology.}
}

In order to see this explicitly, we look qualitatively 
at the effect of the AC model compared to DGP on various
tests of the cosmological expansion history. Several cosmological
datasets are typically used to constrain the expansion history 
at various epochs: type Ia supernovae \cite{DavisWV}, large scale 
structure \cite{BAO}, and the microwave background \cite{WMAP5,WMAPcos}. 
\FIGURE{
\includegraphics[width=16cm]{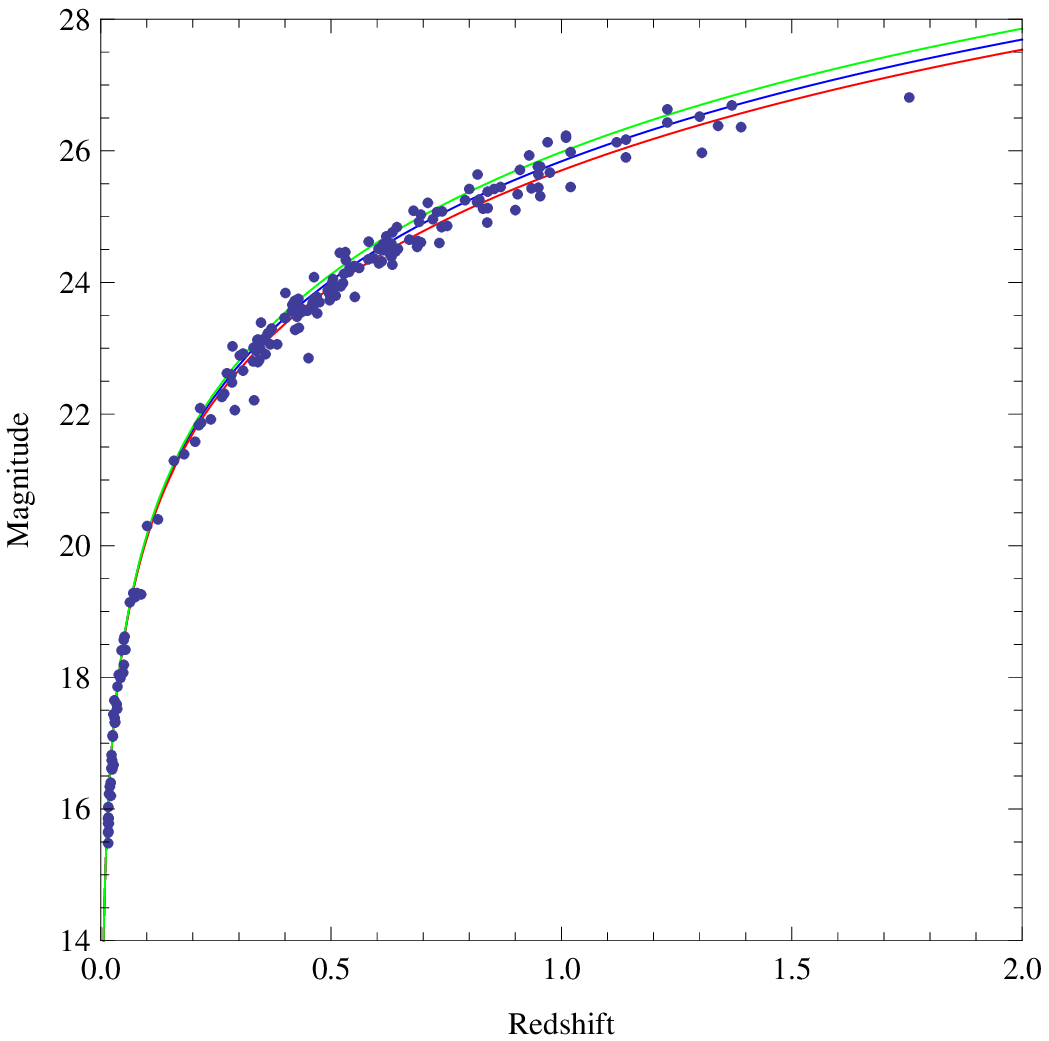}
\caption{Plot of magnitude vs. redshift for (from top to bottom curve)
$\Lambda$CDM (green), DGP (blue) and the AC model (with $\alpha=0.702$, 
$\beta=0.702$) (red),
along with the supernova redshift data. $\Omega_{m}=0.27$.}
\label{fig:magplot}
}

Type Ia supernovae are relatively reliable standard candles, and provide a 
good constraint on the recent expansion of the universe via the
redshift-luminosity relation based on the luminosity distance $d_L$:
\begin{equation}\label{dL}
d_L(z)= \frac{(1+z)}{H_0 \sqrt{|\Omega_k|}} {\cal S} \left ( 
\sqrt{|\Omega_k|} \int_0^z\frac{dz'}{E(z')} \right ) ,
\end{equation}
where ${\cal S}(X) = (X, \sin X, \sinh X)$ for a flat, closed 
or open universe respectively. Since the Hubble parameter is higher
in AC cosmologies (at fixed $\Omega_m$) and increases with increasing
$\beta$, (\ref{dL}) shows that this results in a lessening of the
luminosity distance and hence a lower magnitude.
Figure \ref{fig:magplot} demonstrates this with a direct 
redshift-magnitude plot.

A more conventional visualization of the effect of the AC model is 
given by plotting the preferred regions of $\{\Omega_m,\alpha^2\}$ 
parameter space at different values of $\Omega_k,\;\beta$. 
Figure \ref{fig:snomalph} shows the projection on the $\{\Omega_m,\alpha^2\}$
plane at fixed $\beta$ and fixed $\Omega_k$ values respectively:
$\alpha^2$ is plotted against $\Omega_m$ as this more readily 
compares with the $\Omega_{r_c}$ parameter conventionally used 
in the analysis of DGP models. 
We calculate an expression for $\chi^2$ using the ESSENCE Supernova
dataset \cite{DavisWV}: 
\begin{equation}\label{chi2}
\chi_{red}^2=\frac{1}{192}\sum\frac{(\mu_{obs}-5Log_{10}d_L(z)-43.3)^2}
{\mu_{err}^2}
\end{equation}
where we calculate $d_L(z)$ from (\ref{dL}), using (\ref{asycos}) 
for $E(z)$. The contours plotted in figure \ref{fig:snomalph} 
are then 1$\sigma$ contours, found using $\chi_{min}^2+\Delta\chi^2$, 
where $\Delta\chi^2$ is a standard value and we minimise (\ref{chi2}) 
over the model parameters $\alpha$ and $\Omega_m$.

The left figure indicates how the 
preferred region of parameter space reacts to the $\beta$
parameter (two values, $\beta = 0, \alpha/2$ are shown), and the right
figure how the parameter space reacts to $\Omega_k$ for general $\beta$. In
each case the figure shows that $\alpha$ increases
in response to increasing $\beta$, and although the effect on $\Omega_m$ is less
marked, it decreases slightly. Inspection of (\ref{asycos})
shows why this is so. In effect, fitting the data is accomplished
by keeping $E(z)$ relatively unchanged. If we increase $\beta$, we can
see that this can be mostly counterbalanced by an increase in $\alpha$,
with possible sub-leading changes in the other parameters.
It is interesting to note
that the projection at fixed $\Omega_k$ is relatively insensitive to that 
value of $\Omega_k$, as can be seen from the large overlap 
between the bands, despite the rather large value of $\Omega_k$ chosen
for clarity of the plot. The flat universe band is much broader than
the $\Omega_k=0.1$ band because the effect of the geometry (via the
sinh function in the luminosity distance (\ref{dL})) tends
to magnify any variations in the comoving distance due to variations in
the parameters.

It is clearly
not difficult to reproduce the supernova redshift luminosity relation in
isolation, particularly if the possibility of an open universe is included.
However, the real tension for DGP (and even more so for the asymmetric
model) is in combining the supernova constraints with the constraints from
other cosmological data \cite{DGPcomp}. 
\FIGURE{\label{fig:snomalph}
\includegraphics[width=7cm]{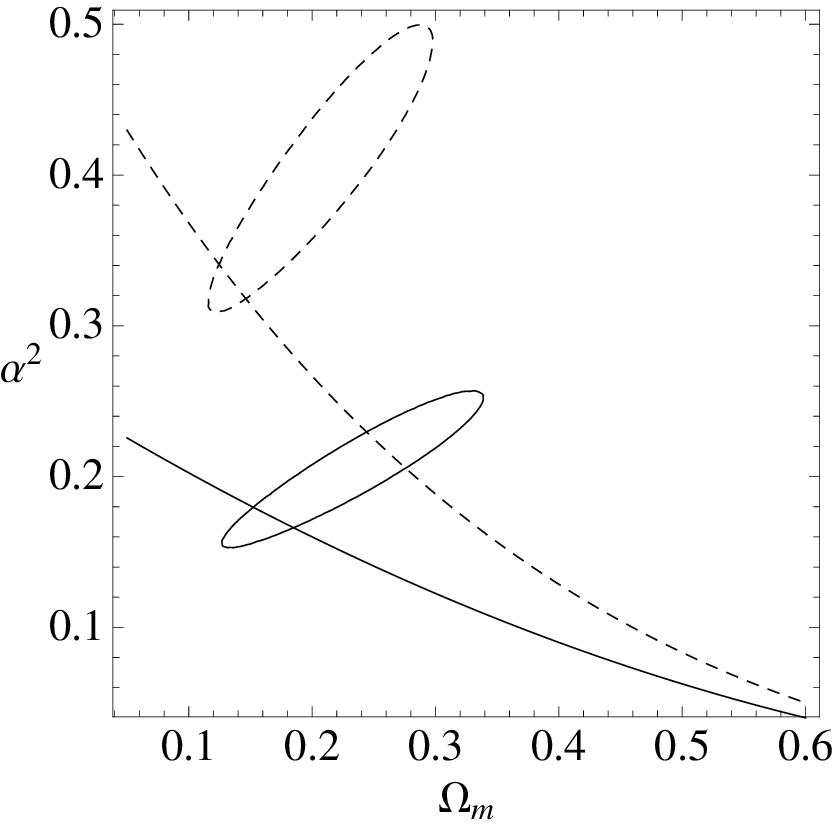}\hskip 5mm\nobreak
\includegraphics[width=7cm]{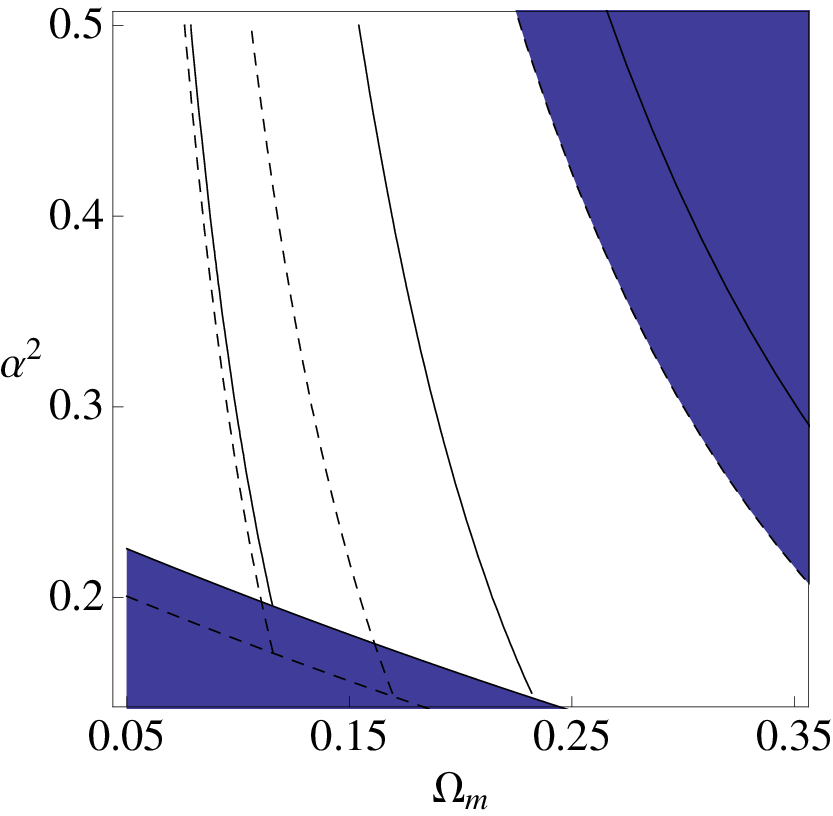}
\caption{An illustration of the constraints on parameter space 
due to the Supernova data. 1$\sigma$ contours are plotted in all cases.
On the left, two fixed values of $\beta$ are shown: $\beta=0$ 
is the solid contour, 
and $\beta=\alpha/2$ the dashed contour (the separate 
lines indicate $\Omega_k=0$ for each $\beta$ value).
In the righthand figure, $\beta$ now varies freely and the bands
indicate two fixed values of $\Omega_k$: the dashed contour, 
$\Omega_k=0.1$, and the solid contour $\Omega_k=0$.
The upper and lower bounding values of $\beta$ are also shown.
}}

The CMB shift parameter \cite{BET}, or (essentially) the ratio between 
the angular diameter distance to and horizon size at decoupling is 
typically used to constrain dark energy models \cite{WMAPcos}, as it 
is relatively model independent:
\be
R(z_*) = \frac{\sqrt{\Omega_m}H_0}{c} (1+z_*) D_A(z_*)
= \frac{\sqrt{\Omega_m}}{\sqrt{|\Omega_k|}} {\cal S} \left ( 
\sqrt{|\Omega_k|} \int_0^{z_*} \frac{dz'}{E(z')} \right ) 
\ee
where $z_* = 1090.51 \pm 0.95$ is the redshift at decoupling \cite{WMAP5}
(and $c$ has been temporarily reintroduced for reference).
The problem with lowering $\Omega_m$ now becomes more apparent. While
we can ensure that the comoving distance is maintained by dropping
$\Omega_m$, the shift parameter is also lowered by this process. Indeed,
flat DGP requires $\Omega_m \simeq 0.35$ to match the WMAP 5 year
value $R = 1.71 \pm 0.02$ \cite{WMAP5,WMAPcos}.
However, one feature which is revealed by the $\beta$ parameter is that
for sufficiently large $\Omega_k$ (or small $\Omega_m$) the angular 
diameter actually increases sufficiently with decreasing $\Omega_m$ to 
outweight this effect, and the shift parameter thus increases with
decreasing $\Omega_m$.

In order to compare the shift parameter constraint on the AC
model to the situation with the DGP model \cite{DGPcomp}, we allow for
open, flat and closed cosmologies, and test the parameter space
compatible with the given shift parameter using the stated WMAP range above.
Figure \ref{fig:shift} shows allowed regions of $\{\Omega_m,\alpha^2\}$ 
parameter space projecting onto fixed $\beta$ and fixed $\Omega_k$ subspaces.

The first figure in Fig.\ \ref{fig:shift} shows three different 
$\beta$ values ranging from the DGP to the Cardassian limit.
These show that as $\beta$ is increased, preferred values of $\Omega_m$ 
become higher. 
The second figure shows three different $\Omega_k$ values. We chose
three indicative values of $\Omega_k$, the flat universe, $\Omega_k = 0.03$
(the best fit value for the open DGP model according 
to the analysis of Song, Sawicki and Hu in \cite{DGPcomp})
and $0.06$ to illustrate the appearance of a minimum in the allowed
region. The minimum appears because decreasing $\Omega_m$ decreases
$E(z)$ over the redshift range, which, together with the magnifying effect
of the sinh function, overwhelms the decrease in the prefactor of
the shift parameter and results in an overall increase of $R_*$.
In this plot, the limiting values of $\beta$ are shown, 
and increasing $\beta$ corresponds
to moving roughly diagonally upwards across the plot. Once again, this
indicates that the preferred value of $\Omega_m$ generally
increases as $\beta$ is increased.
\FIGURE{\label{fig:shift}
\includegraphics[width=7cm]{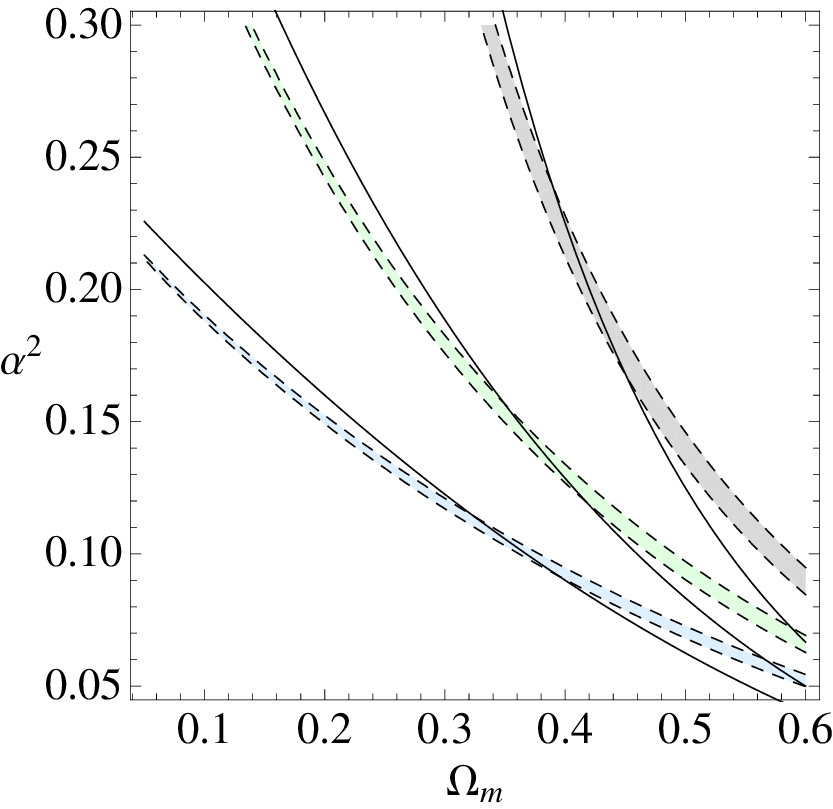}\hskip 5mm\nobreak
\includegraphics[width=7cm]{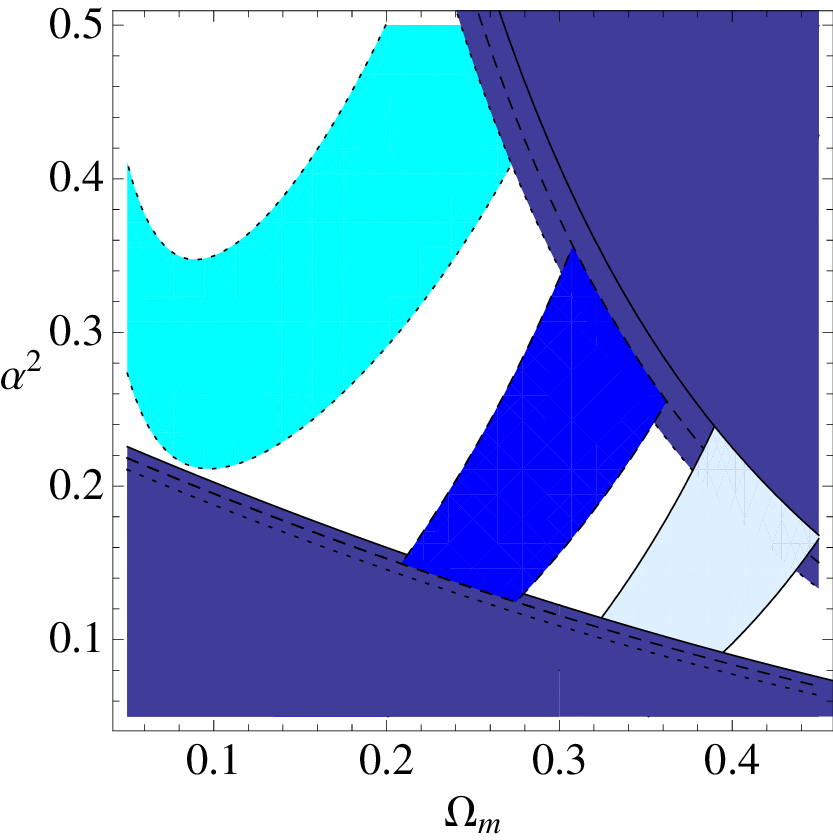}
\caption{A depiction of the region of asymmetric cosmology parameter
space consistent with the shift parameter.
On the left, $\Omega_k$ varies freely, and three fixed values of $\beta$
are shown: $\beta=0$, or the DGP limit, is the lowest (blue) band, 
the green band an intermediate value of $\beta$, and the grey band 
the maximal value of $\beta$. The solid lines indicate $\Omega_k=0$.
On the right, $\beta$ now varies freely and the bands
indicate three fixed values of $\Omega_k$: the lightest band on the
right is a flat universe, the middle dark band is $\Omega_k=0.03$,
and the left hand (cyan) band is $\Omega_k = 0.06$, with the bounding
values of $\beta$ indicated in each case. 
The allowed regions in the plots are obtained using WMAP 5 year values.}
}

We can now see how even just these two constraints on parameter space are
problematic by combining them, since increasing $\beta$ tends to
prefer a decreased $\Omega_m$ to fit the supernova data, yet an
increased $\Omega_m$ to fit the CMB shift parameter.
Increasing $\Omega_k$ in general causes the two ranges for
$\Omega_m$ and $\alpha$ to move together, but unfortunately this pushes
$\Omega_m$ to an unacceptably small value.
In figure \ref{fig:total} we combine the plots, and include for
reference an indication of the constraint coming from the baryon 
acoustic oscillation peak detected by the SDSS survey \cite{BAO}.
This is usually represented as a dimensionless constant
\be
A = D_V(z_1) \frac{\sqrt{\Omega_m} H_0}{cz_1}
= \frac{\sqrt{\Omega_m}}{E(z_1)^{1/3}} \left [ \frac{1}{z_1\sqrt{|\Omega_k|}} 
{\cal S} \left ( \sqrt{|\Omega_k|} \int_0^{z_1} \frac{dz'}{E(z')} \right )
\right]^{2/3} = 0.469\pm0.017
\ee
where $z_1=0.35$, and $D_V$ is the geometric average dilation 
scale \cite{BAO}. There is some debate as to whether this measure 
should be used for models which do not behave as a constant equation 
of state dark energy \cite{BAOdeb}, and in particular where growth of
perturbations may differ significantly from the Einstein case. The
perturbation theory as presented in the appendix indicates
that the AC model is much the same as DGP, and we do 
not propose to add to this debate here. Nonetheless, we include 
this band of parameter space as it is indicative of how serious a
problem structure formation presents.
We do not however use it to constrain our parameters.
For example, in the $\Omega_k = 0, 0.03$ plots, the BAO strip suggests
that while these models might be a significantly poorer fit to the data
than $\Lambda$CDM, structure formation is not a particular problem;
the main issue in these plots is the way the other bands respond to
increasing $\beta$.
\FIGURE{\label{fig:total}
\includegraphics[width=7cm]{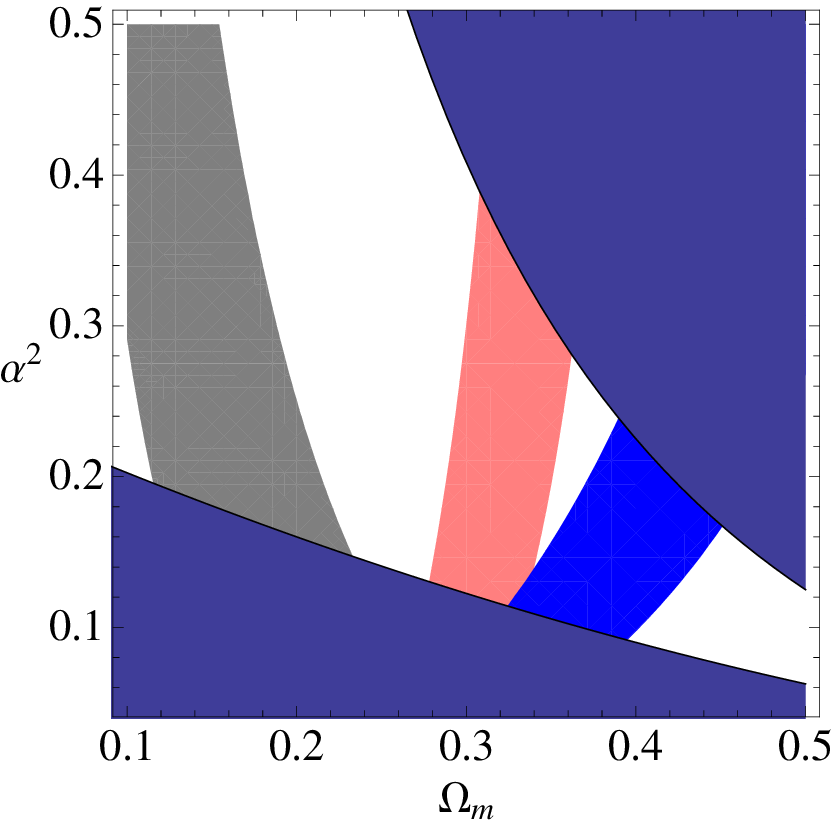}\hskip 5mm\nobreak
\includegraphics[width=7cm]{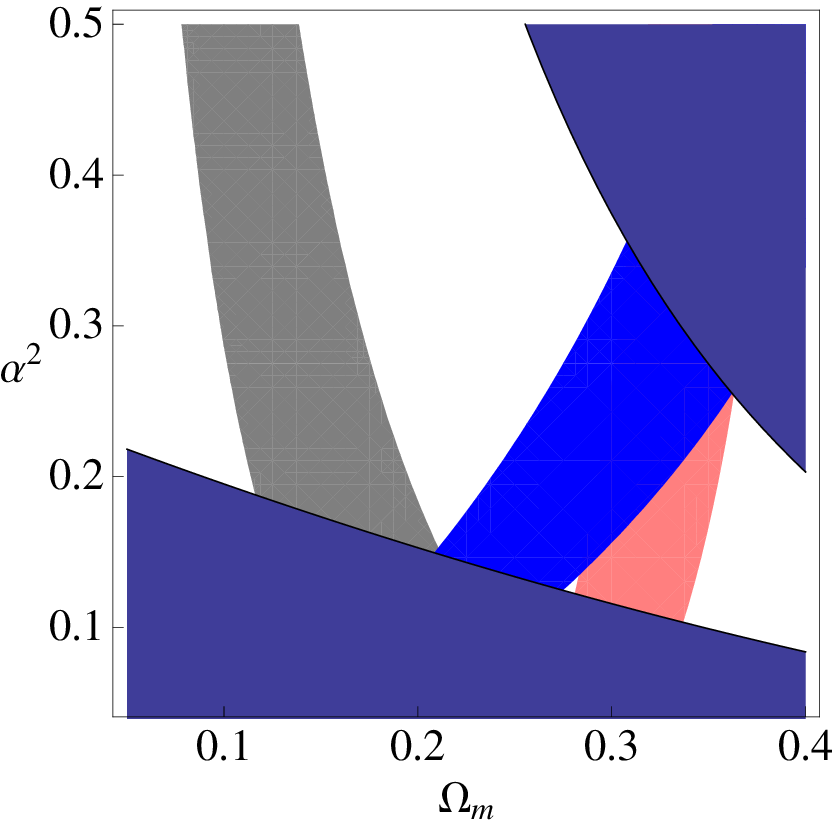}\hskip 5mm\nobreak
\includegraphics[width=7cm]{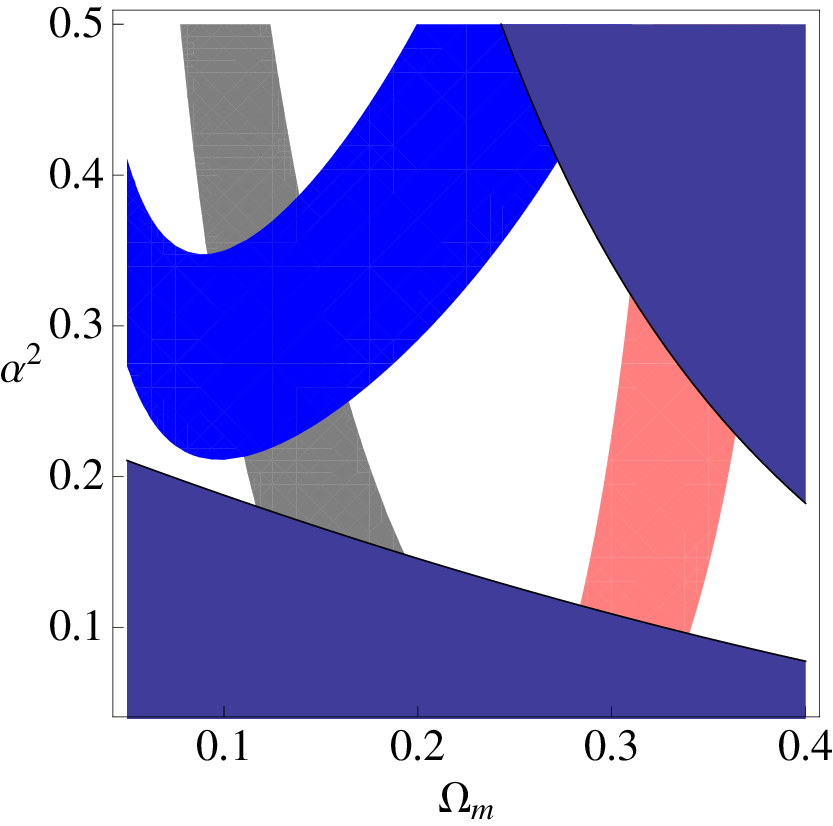}
\caption{A look at the combined effect on the parameter space of
asymmetric cosmology fixing $\Omega_k$ ($0$, $0.03$, and $0.06$ 
respectively) and allowing $\beta$ to vary between
its two limits as indicated.
The dark (blue) band shows the shift parameter preferred
range of $\Omega_m$ and $\alpha^2$, the lighter (grey) band 
from the supernova data, and the lightest (pink) band that from
the BAO constraint. 
} }

Figure \ref{fig:total} shows explicitly the problem of
increasing $\beta$ on the parameter space. For $\Omega_k = 0, 0.03$,
increasing $\beta$ causes the allowed regions of parameter space to
diverge. For the relatively large value of $\Omega_k=0.06$, the SN
and shift parameter regions do overlap, centered around $\alpha^2 \sim 0.28$,
$\Omega_m \sim 0.13$ (with $\beta \sim 0.4\alpha$). However, even 
ignoring the discrepancy with the BAO strip (which is severe) this overlap
occurs at an unacceptably low value of $\Omega_m$. WMAP constrains
$\Omega_m h^2 = 0.1326 \pm 0.0063$ \cite{WMAP5},
meanwhile the Hubble Key Project finds $h = 0.72 \pm 0.08$ 
\cite{HKP}. Therefore, taking the largest allowable value
of $\Omega_m$ as indicated in the final plot of figure \ref{fig:total},
$\Omega_m \simeq 0.17$, requires an $h$ value of $0.86-0.9$, well above
the Hubble Key Project range.

Thus our results show that increasing $\beta$ increases the tension
in fitting the data relative to the DGP model.
In spite of the enlarged parameter space, the asymmetric cosmology
turns out to be under more cosmological tension. This is
because for a given matter content, (\ref{asycos}) shows that
the Hubble parameter increases with redshift more rapidly than in DGP
(which itself is more rapid than $\Lambda$CDM) as the $\beta$ parameter
increases.  This means that for a given $\Omega_m$, the comoving
distance out to a particular redshift is lower in asymmetric gravity
than DGP, which is correspondingly lower than $\Lambda$CDM.
Unfortunately, this means that AC cosmologies are therefore
not a good description of our universe.

\section{Discussion and Model Extensions}

There are other parameters one could include in both DGP and
asymmetric cosmologies. The general bulk spacetime of a cosmological
braneworld includes a bulk black hole \cite{BCG,BCOS}, and while the
effect of this black hole has been considered for Randall-Sundrum
cosmologies (where it gives rise to a dark radiation term) it has
not generally been included in DGP cosmologies (though see \cite{GKMP} for a
discussion of the problems it gives rise to for DGP in general). Adding 
in this general mass term as in (\ref{hgen}) alters (\ref{asyfrweq}) to
\be
\rho \simeq 
3 \frac{M_1^3}{k_1} \left ( H^2 - \frac{\mu_1}{a^4} \right )
- 6 M_2^3 \Big(\sqrt{H^2+k_2^2 - \frac{\mu_2}{a^4} } -k_2\Big)
\ee
where $\mu_1$ is the mass of the bulk black holes in the adS
{\it interior} to the left of the brane, and $\mu_2$ an (effective)
black hole mass of the {\it exterior} adS bulk on the RHS of the brane.

Clearly, having a bulk black hole on the LHS ($\mu_1>0$) simply adds in 
a `dark radiation' term in the effective cosmological energy density in 
an analogous fashion to Randall-Sundrum cosmology \cite{Padillovicz}.
However, the effect of a black hole term on the RHS of the bulk
is more interesting. Since the part of the bulk being excised on 
the RHS is the {\it interior}, we can have either sign for $\mu_2$
(see \cite{GKMP} for potential consistency problems with $\mu_2<0$), 
further, a positive mass bulk black hole actually leads to a 
negative contribution to the brane energy density. Setting $\mu_1=0$,
and writing $\Omega_\mu = \mu_2/(H_0^2a_0^4)$ we find that the 
effective Friedmann equation is only subtly altered in the 
additional braneworld term:
\begin{equation}\label{asycosbh}
E(z)^2 = \Omega_i(1+z)^{3(1+w_i)}
+ 2\alpha(\alpha-\beta) + 2\alpha \sqrt{(\alpha-\beta)^2
+\Omega_i(1+z)^{3(1+w_i)} - \Omega_\mu(1+z)^4}. 
\end{equation}
A negative $\Omega_\mu$ (i.e.\ a negative black hole mass) simply adds to
the value of $E^2$, and therefore will not assist the model in conforming
to the expansion data. However, a {\it positive} bulk black hole mass
contributes negatively, and therefore reduces the value of $E^2$.
However, in order to prevent pressure singularities on the brane,
we must ensure that $\Omega_\mu < \Omega_r$, and thus the best that
can be achieved by this term is a cancellation of the radiation 
density of the universe in the term under the square root, though not
in the leading Einstein term. While this could lead to interesting
effects in the early universe, these will be sub-leading and in any case
it does not significantly help with fitting the late time expansion
of the universe.

To sum up:
We have examined the asymmetric branes model \cite{Padillovicz,tonyboy}, 
a braneworld theory of modified gravity, with a view to exploring how well
it can explain the late-time acceleration of the universe.  The effective
cosmological expansion above a Hubble distance of order $1mm$ is a 
one-parameter generalization of the DGP model, the effect of 
the extra parameter being to retard the expansion of the Universe
relative to DGP. As such, it turns out that the asymmetric model has 
more problems fitting the cosmological expansion data than DGP. In
addition, recent work on ghosts in the stealth model \cite{KPS} suggests
that the AC model may well not be ghost-free around the accelerating
vacuum, thus our overwhelming conclusion is unfortunately that 
pure AC models are not viable cosmological models for late time
acceleration. Nonetheless, it is important
to check the behaviour of all possible concrete modified 
gravity models available to either identify or rule out alternatives 
to $\Lambda$CDM.

\section*{Acknowledgements}
We would like to thank Christos Charmousis, Anwar Gaungoo, 
and Antonio Padilla for useful discussions.
EOC is supported by the European
Commission's Framework Programme 6, through the Marie Curie Early Stage
Training project MEST-CT-2005-021074, and AP is grateful to the 
University of Nottingham for financial support.

\appendix

\section{Perturbation Theory and the Planck Mass}
\label{appA}

Although the asymmetric model naturally lends itself to a boundary description
of the equations of motion, in deriving the Planck mass it is useful to
consider perturbation theory around the ``domain wall" description, 
i.e.\ in which
we take the full range of the coordinate $y$, and represent the brane
as a physical delta function source:
\be
M_i^3 \left ( R_{ab} - \frac{1}{2} R g_{ab} \right ) = 
-\Lambda_i g_{ab} + \delta(y)  \delta^\mu_a \delta^\nu _b \left(
T_{\mu\nu} - \sigma \gamma_{\mu\nu} \right )
\label{deltaeqns}
\ee
Perturbing these equations around the background (\ref{bkmet}) yields:
\bea
M_i^3 \Big [ a^{-2} \partial^2 h_{\mu\nu} 
+ a^{-2} \left [ a^4 (a^{-2}h_{\mu\nu})'
\right]' - 2  a^{-2} \partial_{(\mu}\partial^\lambda {\bar h}_{\nu)\lambda}
&&\nonumber \\
-aa' \left ( a^{-2}h^\lambda_\lambda\right )' \eta_{\mu\nu} \Big ] 
&=& -2\delta(y) [ T_{\mu\nu} - \frac{T}{3} \eta_{\mu\nu} ] \label{teq}\\
\left ( a^{-2}h^\lambda_\lambda\right )'_{,\mu} - \left ( 
a^{-2} \partial^\lambda h_{\mu\lambda} \right ) ' &=& 0 \label{mu5eq}\\
M^3 \left [ a^2 (a^{-2}h^\lambda_\lambda)' \right ] ' &=& -2
\frac{T}{3} \delta(y) \label{55eq} 
\eea

Following the procedure of Garriga and Tanaka \cite{GT}, for constructing the
Green's function,
we see that the bulk solution for the spin 2 mode is 
$h_{\mu\nu} = u_m(y) \chi^{(m)}_{\mu\nu}$, where $\chi$ is
a 4D massive spin 2 tensor, and $u_m$ is found by solving (\ref{teq}):
\be
u_m = A_i J_2\left( \frac{m\zeta_i}{k_i} \right) 
+ B_i N_2\left( \frac{m\zeta_i}{k_i} \right)
\ee
where
\be
\zeta = a^{-1}(z)
\ee
Applying finiteness of the perturbation as $y\to\infty$ implies $B_2=0$.
Meanwhile, continuity and (\ref{teq}) at the brane imply:
\bea
A_1 J_2\left( \frac{m}{k_1} \right) + B_1 N_2\left( \frac{m}{k_1} \right) &=&
A_2 J_2\left( \frac{m}{k_2} \right) \\
M_1^3 \left [ A_1 J_1\left( \frac{m}{k_1} \right) 
+ B_1 N_1\left( \frac{m}{k_1} \right) \right ]&=&
M_2^3 A_2 J_1\left( \frac{m}{k_2} \right) 
\eea
Finally, normalization of the eigenfunctions gives
\be
|A_1|^2 + |B_1|^2 = \frac{m}{k_1} 
\ee
Thus our coefficients are completely specified as:
\bea
A_1 &=& A_2 \; \frac{\pi m}{2k_1} \left [ J_2 \left ( \frac{m}{k_2} \right)
N_1 \left ( \frac{m}{k_1} \right) - \frac{M_2^3}{M_1^3}
J_1 \left( \frac{m}{k_2} \right)
N_2 \left( \frac{m}{k_1} \right) \right ] \\
B_1 &=& - A_2 \frac{\pi m}{2k_1} \left [ J_2 \left ( \frac{m}{k_2} \right)
J_1 \left ( \frac{m}{k_1} \right) - \frac{M_2^3}{M_1^3}
J_1 \left( \frac{m}{k_2} \right)
J_2 \left( \frac{m}{k_1} \right) \right ] 
\eea
where
\bea
A_2^2 &=& \frac{4 k_1}{\pi^2 m} \Big [
J_2^2 \left(\frac{m}{k_2}\right) \left ( N_1^2 \left(\frac{m}{k_1}\right)
+ J_1^2 \left(\frac{m}{k_1}\right) \right ) + 
\frac{M_2^6}{M_1^6} 
J_1^2 \left(\frac{m}{k_2}\right) \left ( N_2^2 \left(\frac{m}{k_1}\right)
+ J_2^2 \left(\frac{m}{k_1}\right) \right ) \nonumber \\
&&- \frac{2M_2^3}{M_1^3} J_1\left(\frac{m}{k_2}\right)
J_2 \left(\frac{m}{k_2}\right) \left ( N_1\left(\frac{m}{k_1}\right)
N_2\left(\frac{m}{k_1}\right) + J_1\left(\frac{m}{k_1}\right)
J_2\left(\frac{m}{k_1}\right) \right ) \Big ] ^{-1}
\eea
Note that, as with the GRS and DGP models, there is no localizable zero mode,
and the spectrum is continuous starting from $m^2=0$. 4D 
gravity must therefore be obtained as an effective behaviour within 
a range of scales. We therefore examine
the Newtonian potential in the brane of a unit mass particle on the brane 
which is given in terms of the eigenfunctions by
\be
V(r) = \frac{2}{M_1^3+M_2^3} \int_0^\infty dm
\frac{e^{-mr}}{4\pi r} |A_2|^2 J_2^2 \left ( \frac{m}{k_2} \right) 
\ee
Then, writing $\varepsilon = M_2^3/M_1^3$, and $r_c = 1/\varepsilon k_1$,
and redefining the integration variable as $x=mr_c/2$
we have to leading order in $\varepsilon$
\be
V(r) \sim \frac{2}{M_1^3} \frac{4 k_1 \varepsilon^2}{4\pi r} \int dx
\frac{x e^{-(2r/r_c)x}}{\left ( 1 - \frac{J_1(2x/k_2r_c) }{xJ_2(2x/k_2r_c)} \right)^2
+\varepsilon^4 x^4}
\ee
In order to estimate this integral, note that
for $r \ll r_c$, this integral is dominated by $x\simeq J_1/J_2 = {\cal O}(1)$,
where the integrand has a value of ${\cal O}(\varepsilon^{-4})$ with a width
of ${\cal O}(\varepsilon^{2})$. Thus
\be
V(r) \propto \frac{k_1}{M_1^3} \; \frac{1}{4\pi r} 
\ee
and the Planck mass can be read off as $m_{pl}^2 \simeq M_1^3/k_1$.


\begin{thebibliography}{}
\bibitem{Riess1} A.~G.~Riess {\it et al.}  
[Supernova Search Team Collaboration],
Astron.\ J.\  {\bf 116}, 1009 (1998) [arXiv:astro-ph/9805201].
A.~G.~Riess {\it et al.}  [Supernova Search Team Collaboration],
Astrophys.\ J.\  {\bf 607}, 665 (2004) [arXiv:astro-ph/0402512].

\bibitem{Perl} S.~Perlmutter {\it et al.}  
[Supernova Cosmology Project Collaboration],
Astrophys.\ J.\  {\bf 517}, 565 (1999) [arXiv:astro-ph/9812133].

\bibitem{WMAP5}
J.~Dunkley {\it et al.}  [WMAP Collaboration],
Astrophys.\ J.\ Suppl.\  {\bf 180}, 306 (2009)
[arXiv:0803.0586 [astro-ph]].
Astrophys.\ J.\ Suppl.\  {\bf 170}, 377 (2007) [arXiv:astro-ph/0603449].
D.~N.~Spergel {\it et al.}  [WMAP Collaboration],
Astrophys.\ J.\ Suppl.\  {\bf 148}, 175 (2003) [arXiv:astro-ph/0302209].

\bibitem{Cole:2005sx} S.~Cole {\it et al.}  [The 2dFGRS Collaboration],
Mon.\ Not.\ Roy.\ Astron.\ Soc.\  {\bf 362}, 505 (2005) 
[arXiv:astro-ph/0501174].
M.~Tegmark {\it et al.}  [SDSS Collaboration],
Phys.\ Rev.\  D {\bf 74}, 123507 (2006) [arXiv:astro-ph/0608632].

\bibitem{quint}
R.~R.~Caldwell, R.~Dave and P.~J.~Steinhardt,
Phys.\ Rev.\ Lett.\  {\bf 80}, 1582 (1998)
[arXiv:astro-ph/9708069].

\bibitem{card}
K.~Freese and M.~Lewis,
Phys.\ Lett.\  B {\bf 540}, 1 (2002)
[arXiv:astro-ph/0201229].

\bibitem{hobbit}
V.~F.~Cardone, A.~Troisi and S.~Capozziello,
Phys.\ Rev.\  D {\bf 69}, 083517 (2004)
[arXiv:astro-ph/0402228].

\bibitem{MOND}
M.~Milgrom,
Astrophys.\ J.\  {\bf 270}, 365 (1983).
J.~D.~Bekenstein,
Phys.\ Rev.\  D {\bf 70}, 083509 (2004)
[Erratum-ibid.\  D {\bf 71}, 069901 (2005)]
[arXiv:astro-ph/0403694].

\bibitem{fofR}
S.~M.~Carroll, V.~Duvvuri, M.~Trodden and M.~S.~Turner,
Phys.\ Rev.\  D {\bf 70}, 043528 (2004)
[arXiv:astro-ph/0306438].
S.~M.~Carroll, A.~De Felice, V.~Duvvuri, D.~A.~Easson, 
M.~Trodden and M.~S.~Turner,
Phys.\ Rev.\  D {\bf 71}, 063513 (2005)
[arXiv:astro-ph/0410031].
D.~A.~Easson,
Int.\ J.\ Mod.\ Phys.\  A {\bf 19}, 5343 (2004)
[arXiv:astro-ph/0411209].

\bibitem{ADD}
N.~Arkani-Hamed, S.~Dimopoulos and G.~R.~Dvali,
Phys.\ Lett.\  B {\bf 429}, 263 (1998) [arXiv:hep-ph/9803315].
N.~Arkani-Hamed, S.~Dimopoulos and G.~R.~Dvali,
Phys.\ Rev.\  D {\bf 59}, 086004 (1999) [arXiv:hep-ph/9807344].
I.~Antoniadis, N.~Arkani-Hamed, S.~Dimopoulos and G.~R.~Dvali,
Phys.\ Lett.\  B {\bf 436}, 257 (1998) [arXiv:hep-ph/9804398].
N.~Arkani-Hamed, S.~Dimopoulos, G.~R.~Dvali and N.~Kaloper,
Phys.\ Rev.\ Lett.\  {\bf 84}, 586 (2000) [arXiv:hep-th/9907209].

\bibitem{RS}
L.~Randall and R.~Sundrum,
Phys.\ Rev.\ Lett.\  {\bf 83}, 3370 (1999) [arXiv:hep-ph/9905221].
L.~Randall and R.~Sundrum,
Phys.\ Rev.\ Lett.\  {\bf 83}, 4690 (1999) [arXiv:hep-th/9906064].

\bibitem{mill}
I.~I.~Kogan, S.~Mouslopoulos, A.~Papazoglou, G.~G.~Ross and J.~Santiago,
Nucl.\ Phys.\  B {\bf 584}, 313 (2000)
[arXiv:hep-ph/9912552].
I.~I.~Kogan and G.~G.~Ross,
Phys.\ Lett.\  B {\bf 485}, 255 (2000)
[arXiv:hep-th/0003074].

\bibitem{GRS}
R.~Gregory, V.~A.~Rubakov and S.~M.~Sibiryakov,
Phys.\ Rev.\ Lett.\  {\bf 84}, 5928 (2000)
[arXiv:hep-th/0002072].
R.~Gregory, V.~A.~Rubakov and S.~M.~Sibiryakov,
Phys.\ Lett.\  B {\bf 489}, 203 (2000)
[arXiv:hep-th/0003045].

\bibitem{DGP}
G.~R.~Dvali, G.~Gabadadze and M.~Porrati,
Phys.\ Lett.\  B {\bf 485}, 208 (2000) [arXiv:hep-th/0005016].

\bibitem{VS}
V.~Sahni and Y.~Shtanov,
JCAP {\bf 0311}, 014 (2003)
[arXiv:astro-ph/0202346].

\bibitem{DGPcos}
C.~Deffayet,
Phys.\ Lett.\  B {\bf 502}, 199 (2001) [arXiv:hep-th/0010186].
C.~Deffayet, G.~R.~Dvali and G.~Gabadadze,
Phys.\ Rev.\  D {\bf 65}, 044023 (2002) [arXiv:astro-ph/0105068].
C.~Deffayet, S.~J.~Landau, J.~Raux, M.~Zaldarriaga and P.~Astier,
Phys.\ Rev.\  D {\bf 66}, 024019 (2002) [arXiv:astro-ph/0201164].

\bibitem{GHOST}
K.~Koyama,
Phys.\ Rev.\  D {\bf 72}, 123511 (2005) [arXiv:hep-th/0503191].
D.~Gorbunov, K.~Koyama and S.~Sibiryakov,
Phys.\ Rev.\  D {\bf 73}, 044016 (2006) [arXiv:hep-th/0512097].
M.~S.~Carena, J.~Lykken, M.~Park and J.~Santiago,
Phys.\ Rev.\  D {\bf 75}, 026009 (2007)
[arXiv:hep-th/0611157].

\bibitem{CGKP}
C.~Charmousis, R.~Gregory, N.~Kaloper and A.~Padilla,
JHEP {\bf 0610}, 066 (2006) [arXiv:hep-th/0604086].

\bibitem{DGPsing}
N.~Kaloper,
Phys.\ Rev.\ Lett.\  {\bf 94}, 181601 (2005)
[Erratum-ibid.\  {\bf 95}, 059901 (2005)]
[arXiv:hep-th/0501028].
N.~Kaloper,
Phys.\ Rev.\  D {\bf 71}, 086003 (2005)
[Erratum-ibid.\  D {\bf 71}, 129905 (2005)]
[arXiv:hep-th/0502035].

\bibitem{GKMP}
K.~Izumi, K.~Koyama, O.~Pujolas and T.~Tanaka,
Phys.\ Rev.\  D {\bf 76}, 104041 (2007)
[arXiv:0706.1980 [hep-th]].
R.~Gregory, N.~Kaloper, R.~C.~Myers and A.~Padilla,
JHEP {\bf 0710}, 069 (2007)
[arXiv:0707.2666 [hep-th]].
R.~Gregory,
Prog.\ Theor.\ Phys.\ Suppl.\  {\bf 172}, 71 (2008)
[arXiv:0801.1603 [hep-th]].

\bibitem{DG}
C.~Deffayet, G.~Gabadadze and A.~Iglesias,
JCAP {\bf 0608}, 012 (2006)
[arXiv:hep-th/0607099].
G.~Dvali,
New J.\ Phys.\  {\bf 8}, 326 (2006)
[arXiv:hep-th/0610013].

\bibitem{Padillovicz}
A.~Padilla,
Class.\ Quant.\ Grav.\  {\bf 22}, 681 (2005) [arXiv:hep-th/0406157].
 
\bibitem{tonyboy}
A.~Padilla,
Class.\ Quant.\ Grav.\  {\bf 22}, 1087 (2005) [arXiv:hep-th/0410033].

\bibitem{CGP}
C.~Charmousis, R.~Gregory and A.~Padilla,
JCAP {\bf 0710}, 006 (2007) [arXiv:0706.0857 [hep-th]].

\bibitem{SSST}
Y.~Shtanov, V.~Sahni, A.~Shafieloo and A.~Toporensky,
``Induced cosmological constant and other features of asymmetric brane
embedding,''
arXiv:0901.3074 [gr-qc].

\bibitem{KK}
K.~Koyama and K.~Koyama,
Phys.\ Rev.\  D {\bf 72}, 043511 (2005)
[arXiv:hep-th/0501232].

\bibitem{DavisWV} 
T.~M.~Davis {\it et al.},
Astrophys.\ J.\  {\bf 666}, 716 (2007) [arXiv:astro-ph/0701510].
W.~M.~Wood-Vasey {\it et al.}  
[ESSENCE Collaboration],
Astrophys.\ J.\  {\bf 666}, 694 (2007) [arXiv:astro-ph/0701041].
A.~G.~Riess {\it et al.},
Astrophys.\ J.\  {\bf 659}, 98 (2007) [arXiv:astro-ph/0611572].

\bibitem{WMAPcos}
E.~Komatsu {\it et al.}  [WMAP Collaboration],
Astrophys.\ J.\ Suppl.\  {\bf 180}, 330 (2009)
[arXiv:0803.0547 [astro-ph]].

\bibitem{EBW}
V.~A.~Rubakov and M.~E.~Shaposhnikov,
Phys.\ Lett.\ B {\bf 125}, 139 (1983).\\
V.~A.~Rubakov and M.~E.~Shaposhnikov,
Phys.\ Lett.\ B {\bf 125}, 136 (1983).\\
K.~Akama,
Lect.\ Notes Phys.\  {\bf 176}, 267 (1982).
[arXiv:hep-th/0001113].

\bibitem{Israel}
W.~Israel,
Nuovo Cim.\  B {\bf 44S10}, 1 (1966) 
[Erratum-ibid.\  B {\bf 48}, 463 (1967\ NUCIA,B44,1.1966)].

\bibitem{GG}
D.~Garfinkle and R.~Gregory,
Phys.\ Rev.\  D {\bf 41}, 1889 (1990).
F.~Bonjour, C.~Charmousis and R.~Gregory,
Phys.\ Rev.\  D {\bf 62}, 083504 (2000)
[arXiv:gr-qc/0002063].


\bibitem{GTV}
D.~P.~George, M.~Trodden and R.~R.~Volkas,
JHEP {\bf 0902}, 035 (2009)
[arXiv:0810.3746 [hep-ph]].

\bibitem{BCG}
P.~Bowcock, C.~Charmousis and R.~Gregory,
Class.\ Quant.\ Grav.\  {\bf 17}, 4745 (2000)
[arXiv:hep-th/0007177].

\bibitem{Kapner} 
D.~J.~Kapner, T.~S.~Cook, E.~G.~Adelberger, J.~H.~Gundlach, 
B.~R.~Heckel, C.~D.~Hoyle and H.~E.~Swanson,
Phys.\ Rev.\ Lett.\  {\bf 98}, 021101 (2007) [arXiv:hep-ph/0611184].
E.~G.~Adelberger, B.~R.~Heckel and A.~E.~Nelson,
Ann.\ Rev.\ Nucl.\ Part.\ Sci.\  {\bf 53}, 77 (2003) [arXiv:hep-ph/0307284].

\bibitem{DGPcomp}
R.~Maartens and E.~Majerotto,
Phys.\ Rev.\  D {\bf 74}, 023004 (2006)
[arXiv:astro-ph/0603353].
Y.~S.~Song, I.~Sawicki and W.~Hu,
Phys.\ Rev.\  D {\bf 75}, 064003 (2007)
[arXiv:astro-ph/0606286].
S.~Rydbeck, M.~Fairbairn and A.~Goobar,
JCAP {\bf 0705}, 003 (2007)
[arXiv:astro-ph/0701495].

\bibitem{Varun}
V.~Sahni,
Lect.\ Notes Phys.\  {\bf 653}, 141 (2004)
[arXiv:astro-ph/0403324].
V.~Sahni and A.~Starobinsky,
Int.\ J.\ Mod.\ Phys.\  D {\bf 15}, 2105 (2006)
[arXiv:astro-ph/0610026].
M.~Ishak, A.~Upadhye and D.~N.~Spergel,
Phys.\ Rev.\  D {\bf 74}, 043513 (2006)
[arXiv:astro-ph/0507184].

\bibitem{BAO}
D.~J.~Eisenstein {\it et al.}  [SDSS Collaboration],
Astrophys.\ J.\  {\bf 633}, 560 (2005)
[arXiv:astro-ph/0501171].
W.~J.~Percival, S.~Cole, D.~J.~Eisenstein, R.~C.~Nichol, 
J.~A.~Peacock, A.~C.~Pope and A.~S.~Szalay,
Mon.\ Not.\ Roy.\ Astron.\ Soc.\  {\bf 381}, 1053 (2007)
[arXiv:0705.3323 [astro-ph]].

\bibitem{BET}
J.~R.~Bond, G.~Efstathiou and M.~Tegmark,
Mon.\ Not.\ Roy.\ Astron.\ Soc.\  {\bf 291}, L33 (1997)
[arXiv:astro-ph/9702100].

\bibitem{BAOdeb}
J.~Dick, L.~Knox and M.~Chu,
JCAP {\bf 0607}, 001 (2006)
[arXiv:astro-ph/0603247].
V.~Barger, Y.~Gao and D.~Marfatia,
Phys.\ Lett.\  B {\bf 648}, 127 (2007)
[arXiv:astro-ph/0611775].
Y.~Wang,
``Clarifying Forecasts of Dark Energy Constraints from Baryon Acoustic
Oscillations,''
arXiv:0904.2218 [astro-ph.CO].

\bibitem{HKP}
W.~L.~Freedman {\it et al.}  [HST Collaboration],
Astrophys.\ J.\  {\bf 553}, 47 (2001)
[arXiv:astro-ph/0012376].

\bibitem{BCOS}
H.~A.~Chamblin and H.~S.~Reall,
Nucl.\ Phys.\  B {\bf 562}, 133 (1999)
[arXiv:hep-th/9903225].
P.~Binetruy, C.~Deffayet and D.~Langlois,
Nucl.\ Phys.\ B {\bf 565}, 269 (2000)
[arXiv:hep-th/9905012].
N.~Kaloper,
Phys.\ Rev.\ D {\bf 60}, 123506 (1999)
[arXiv:hep-th/9905210].
P.~Kraus,
JHEP {\bf 9912}, 011 (1999)
[arXiv:hep-th/9910149].

\bibitem{KPS}
K.~Koyama, A.~Padilla and F.~P.~Silva,
``Ghosts in asymmetric brane gravity and the decoupled stealth limit,''
arXiv:0901.0713 [hep-th].

\bibitem{GT}
J.~Garriga and T.~Tanaka,
Phys.\ Rev.\ Lett.\  {\bf 84}, 2778 (2000)
[arXiv:hep-th/9911055].

\end{thebibliography}
\end{document}